\documentclass[10pt, conference]{IEEEtran}
\IEEEoverridecommandlockouts
\usepackage[
  backend=biber,
  maxcitenames=3,
  minbibnames=1,
  mincitenames=1,
  maxbibnames=3,
  uniquelist=false,
  style=ieee,
  url=false,
  doi=false,
  eprint=false,
  isbn=false,
]{biblatex}
\AtBeginBibliography{\footnotesize}
\usepackage{booktabs}
\usepackage{csquotes}
\usepackage{tablefootnote}
\usepackage{numprint}
\usepackage{nicematrix}
\usepackage[compat=0.6]{yquant}
\usepackage{lipsum}
\usepackage{comment}
\usepackage{tikz}
\usetikzlibrary{arrows.meta}
\usetikzlibrary{positioning,fit}
\usetikzlibrary{patterns,calc}
\usetikzlibrary{tikzmark}
\usepackage{fp}
\usepackage{xspace}
\usepackage{amsmath,amssymb,amsfonts}
\usepackage{bbm}
\usepackage{quantikz}
\usepackage{placeins}
\usepackage[T1]{fontenc}
\usepackage[colorlinks=true,urlcolor=teal,
            linkcolor=teal,citecolor=teal]{hyperref}
\usepackage[most]{tcolorbox}
\usepackage{listings}
\usepackage{censor}
\definecolor{lfd1}{HTML}{000000}
\definecolor{lfd2}{HTML}{E69F00}
\definecolor{lfd3}{HTML}{999999}
\definecolor{lfd4}{HTML}{009371}
\definecolor{lfd5}{HTML}{beaed4}
\definecolor{lfd6}{HTML}{ed665a}
\definecolor{lfd7}{HTML}{1f78b4}
\definecolor{bggray}{gray}{0.9}

\newcommand{\reprolink}{https://github.com/lfd/RL_for_JO}
\newcommand{\ie}{\emph{i.e.}\xspace}
\newcommand{\eg}{\emph{e.g.}\xspace}
\newcommand{\etal}{\emph{et al.}\xspace}
\newcommand{\rj}{\emph{ReJoin}\xspace}
\newcommand{\repro}{\href{\reprolink}{reproduction package}\xspace}
\newcommand{\mapa}{Marcus and Papaemmanouil\xspace}

\makeatletter
\def\calcLength(#1,#2)#3{%
\pgfpointdiff{\pgfpointanchor{#1}{center}}%
             {\pgfpointanchor{#2}{center}}%
\pgf@xa=\pgf@x%
\pgf@ya=\pgf@y%
\FPeval\@temp@a{\pgfmath@tonumber{\pgf@xa}}%
\FPeval\@temp@b{\pgfmath@tonumber{\pgf@ya}}%
\FPeval\@temp@sum{(\@temp@a*\@temp@a+\@temp@b*\@temp@b)}%
\FProot{\FPMathLen}{\@temp@sum}{2}%
\FPround\FPMathLen\FPMathLen5\relax
\global\expandafter\edef\csname #3\endcsname{\FPMathLen}
}
\makeatother

\lstdefinestyle{query}{
  language=SQL,
  stepnumber=1,
  numbersep=10pt,
  tabsize=4,
  showspaces=false,
  showstringspaces=false,
  basicstyle=\linespread{1}\fontfamily{lmtt}\selectfont\small,
  keywordstyle=\color{blue},
  stringstyle=\color{purple},
  upquote=true,
  breaklines=true,
  commentstyle=\color{CadetBlue}
}

\definecolor{mygray}{rgb}{0.643,0.643,0.643}
\newtcolorbox{querybox}[2][]{%
  sidebyside align=top,
  enhanced,
  boxsep=2pt,
  arc=0pt,
  top=-3pt, bottom=-3pt,
  left=2pt, right=0pt,
  colback=white,
  colframe=mygray,
  boxrule=0.5pt,
  leftrule=12pt,
  overlay unbroken and first ={%
    \node[rotate=90,
          minimum width=0.5cm,
          anchor=south,
          yshift=-11pt,
          white]
    at (frame.west) {#2};
  }
}

\newtcolorbox{matrixbox}[2][]{%
  sidebyside align=top,
  enhanced,
  boxsep=0pt,
  arc=0pt,
  left=-1em,
  top=-0.8em,
  boxrule=0pt,
  colframe=bggray,
  colback=bggray,
  leftrule=12pt,
  overlay unbroken and first ={%
    \node[rotate=90,
          minimum width=0.5cm,
          anchor=south west,
          font=\itshape,
          yshift=0pt, 
          xshift=0.5em,
          black]
    at (frame.south west) {#2};
  }
}

\newcommand{\CZgate}{\ensuremath{\text{C\raisebox{0.08em}{--}}\!\hat{Z}}\xspace}

\newcommand{\nop}[1]{}

\newlength{\WIDTH}\newlength{\HEIGHT}
\renewcommand{\Join}{\bowtie}

\newboolean{anonymous}
\setboolean{anonymous}{false}
\ifbool{anonymous}{}{\renewcommand{\censor}[1]{#1}}
\ifbool{anonymous}{}{\renewcommand{\blackout}[1]{#1}}
\ifbool{anonymous}{\newcommand{\genemail}[2]{\href{mailto:xxx.xxx@xx.xx}{\blackout{#2}}}}{\newcommand{\genemail}[2]{\href{#1}{#2}}}

\begin{document}
\title{Hype or Heuristic? Quantum Reinforcement Learning for Join Order Optimisation}

\author{\IEEEauthorblockN{\blackout{Maja Franz}}
\IEEEauthorblockA{\blackout{\textit{Technical University of}}\\
    \blackout{\textit{Applied Sciences Regensburg}} \\
    \blackout{Regensburg, Germany} \\
    \genemail{mailto:maja.franz@othr.de}{maja.franz@othr.de}}
\and
\IEEEauthorblockN{\blackout{Tobias Winker}}
\IEEEauthorblockA{\blackout{\textit{University of Lübeck}}\\
    \blackout{Lübeck, Germany}\\
    \genemail{mailto:t.winker@uni-luebeck.de}{t.winker@uni-luebeck.de}}
\and
\IEEEauthorblockN{\blackout{Sven Groppe}}
\IEEEauthorblockA{\blackout{\textit{University of Lübeck}}\\
    \blackout{Lübeck, Germany}\\
    \genemail{mailto:sven.groppe@uni-luebeck.de}{sven.groppe@uni-luebeck.de}}
\and
\IEEEauthorblockN{\blackout{Wolfgang Mauerer}}
\IEEEauthorblockA{\blackout{\textit{Technical University of}}\\
    \blackout{\textit{Applied Sciences Regensburg}}\\
    \blackout{\textit{Siemens AG, Technology}}\\
    \blackout{Regensburg/Munich, Germany}\\
    \genemail{mailto:wolfgang.mauerer@othr.de}{wolfgang.mauerer@othr.de}}
}

\maketitle

\begin{abstract}
Identifying optimal join orders (JOs) stands out as a key challenge in database research and engineering.
Owing to the large search space, established classical methods rely on approximations and heuristics.
Recent efforts have successfully explored reinforcement learning (RL) for JO. Likewise, quantum versions of RL have received considerable scientific 
attention. Yet, it is an open question if they can achieve
sustainable, overall practical advantages with improved quantum processors.

In this paper, we present a novel approach that uses quantum reinforcement learning (QRL) 
for JO based on a hybrid variational quantum ansatz.
It is able to handle general \emph{bushy} join trees instead of resorting to
simpler \emph{left-deep} variants as compared to approaches based on quantum(-inspired) optimisation,
yet requires multiple orders of magnitudes fewer qubits, which
is a scarce resource even for post-NISQ systems.

Despite moderate circuit depth, the ansatz exceeds current NISQ capabilities,
which requires an evaluation by numerical simulations. While QRL may not 
significantly outperform classical approaches in solving the JO problem with
respect to result quality (albeit we see parity), we find a drastic reduction in required trainable parameters. This benefits practically relevant aspects ranging 
from shorter training times compared to classical RL, less involved classical optimisation passes, or better use of available training data,
and fits data-stream and low-latency processing scenarios.
Our comprehensive evaluation and careful discussion delivers a balanced 
perspective on possible practical quantum advantage, provides
insights for future systemic approaches, and allows for quantitatively
assessing trade-offs of quantum approaches for one of the most crucial
problems of database management systems.
\end{abstract}


\section{Introduction}
\label{sec:intro}
In database research and industrial practice, finding
good orders in which joins between table columns are
executed in a query---the so-called
join order (JO) problem---counts amongs the most fundamental issues
of database management systems (DMBS)~\cite{steinbrunn97, neumann09, neumann18, trummer17, han09, kolchinsky18, goncalves14, Leis.2018, Moerkotte.2020}. The chosen order
substantially impacts query execution time.
While the problem does only need little amounts of input information (the query
to be executed, and characteristics of the payload data 
obtained from statistical samples), the problem is know to be NP-hard in
general, and also for common restricted scenarios~\cite{cluet95}. An optimal JO cannot be efficiently found deterministically. The last few decades have seen various classical heuristics that can find suboptimal JOs in polynomial time~\cite{swami89, krishnamurthy86, selinger79}.
Recent classical work~\cite{marcus18, marcus19, krishnan18, trummer21, chen22, xiang20, wang2023, ji2023} explores the application of reinforcement learning (RL) to tackle the JO problem.
RL is considered to be beneficial in scenarios where the solution to a problem can be
determined by a series of subsequent decision steps, and where finding one such good sequence for a problem generalises well to others, or when highly dynamic problems are
considered.  By learning from experience of past query evaluation, RL can find good
decision sequences in vast search spaces, and only requires information about the current 
state of the system. This is particularly advantageous for the JO problem,
as very typical scenarios in database systems need to process information at 
a high temporal frequency.

In this paper, we approach RL for JO from the perspective of quantum machine learning (QML), an emerging technique that leverages the principles of quantum mechanics for potential computational speed-ups.
It has been shown that certain problems~\cite{shor99, grover96} can be solved more efficiently using quantum algorithms over classical approaches.
However, the practical utility of these algorithms is limited on the current generation of quantum computers, so-called noisy intermediate-scale quantum (NISQ) systems~\cite{preskill18}, as they only offer a limited amount of qubits and are prone to noise and imperfections~\cite{greiwe23} that strongly limit possible circuits depth and thus the length of quantum computations.
To address these limitations, \emph{hybrid} quantum-classical algorithms are proposed, where only a limited number of steps is performed on a quantum computer and the remaining steps on classical machines.
As Pirnay~\etal~\cite{pirnay23} show, fault-tolerant quantum computers can provably
provide super-polynomial advantage for optimisation problems over classical algorithms. 
Hybrid \emph{variational} algorithms~\cite{farhi14, mclean16, cerezo21} are considered 
key candidates for exploiting advantages of near-term quantum devices, but
could also be beneficial in post-NISQ systems because of their resource efficiency.

Within the class of hybrid variational algorithms, quantum machine learning (QML) has shown promise by moving certain parts of classical machine learning to quantum computers.
QCs will, despite common misperceptions, likely be inapt for handling large amounts of data~\cite{hoefler23}. This makes quantum reinforcement learning (QRL)~\cite{chen20, skolik22, franz22}, which requires little training data, a promising approach.
As established JO approaches mostly rely on statistical estimates
of properties of the database, JO seems a good match for QRL.

QML in general has been shown to outperform classical machine learning for certain tasks~\cite{dilip22, huang22, du20, huang21, liu21, sweke21, havl19}.
Specifically, for QRL it is hypothesised that fewer parameters are required than for classical neural networks (NNs) to address RL tasks~\cite{lockwood20, chen20}.
Several studies also suggest that QRL can solve tasks that are intractable to classical machine learning~\cite{jerbi21}, or that it may have an advantage over classical NNs in terms of sampling complexity, that is, fewer interactions with the environment are required to achieve optimality for certain problems~\cite{skolik22, franz22}.
For these reasons, the application of QML to database problems is also considered promising~\cite{calikyilmaz23}.

However, as detailed in Ref.~\cite{schuld22}, many approaches for QML that claim quantum advantage rest on artificially constructed scenarios (\eg, \cite{havl19, huang21, liu21}).
Consequently, a practical definition of QML goals is required, which should not imply an exponential speed-up compared to classical approaches, but rather is a matter of details.

We have chosen to use a recent classical
RL-based approach to join ordering by \mapa~\cite{marcus18} as
baseline that is well aligned with intensively studied quantum variants
of reinforcement learning~\cite{meyer22}.
It is known that a careful consideration of various factors is necessary to gauge
potential improvements.
This includes a sound classical baseline, data representation, quantum circuit structure, and hyperparameters.
Further, we provide a high-level evaluation of hardware requirements. Our detailed contributions are:

\begin{itemize}
  \item We systematically replicate\footnote{We follow
      ACM terminology on \href{https://www.acm.org/publications/policies/artifact-review-and-badging-current}{Artifact Review and Badging}: A 
      \emph{replication} describes measurements obtained by a different team 
      using a different experimental setup. The term \emph{re-implementation} is 
      also common in the literature, with identical meaning.} the classical baseline~\cite{marcus18} and generalise it to the quantum case. As the baseline does not provide
      source code or hyperparameters, this is an important prerequisite to ascertain a fair comparison, and allows us to consider all aspects of the DBMS.
  \item We comprehensively simulate the performance of our approach on the join order benchmark (JOB), which is
  a universally accepted touchstone in the database community, and compare it against the classical baseline and a single-step QML technique~\cite{winker23} that was shown to outperform established classical approaches.
    Multi-step QRL can achieve up to 17\% lower median costs than single-step QML on the selected dataset and cost model.
  \item We identify potentials for improvement in view of future hardware development, and 
  carefully address the issue of judging realistic potentials for practical improvements over classical heuristics.
  \item We provide an open-source \repro~\cite{mauerer:22:q-saner} that makes our code
  transparent to the community, and can serve as basis to build further
  experiments upon, and benchmark alternative approaches against.
\end{itemize}

We aim to provide a comprehensive perspective on the quantum advantage landscape in RL for the JO problem.
By combining optimistic hypotheses with an acknowledgement of established challenges and limitations, we strive to present a balanced view.
This balance is important to guiding future research directions and manage expectations regarding the (near- and far-term) practical benefits of quantum algorithms in the
field of database management systems.

The paper is structured as follows:
\autoref{sec:rel-work} reviews existing literature on classical approaches for the JO problem and QC for databases.
\autoref{sec:preliminaries} describes the theoretical background for the application of the JO problem and the method of classical and quantum RL, followed by an overview of our methodology in \autoref{sec:methodology}.
\autoref{sec:exp} outlines our experiments, which are discussed in \autoref{sec:discussion}.
We conclude in \autoref{sec:concl}.

\section{Related Work}\label{sec:rel-work}

The problem of query optimisation, which is formally defined in \autoref{sec:bg-jo}, has been studied for over 40 
years~\cite{selinger79}, and new results appear frequently~\cite{Leis.2018, Moerkotte.2020, neumann18, trummer17}.
Since the search space for the JO problem scales factorial~\cite{cluet95}, an exhaustive search for the optimal JO is only feasible for a small number of relations, even when relying on dynamic programming (DP) approaches~\cite{meister20, moerkotte06, selinger79, vance96, moerkotte08}, necessitating heuristic methods~\cite{steinbrunn97, horng94, bruno10, ioannidis90, swami89, trummer16a} for large queries.

Heuristics require to calculate costs; for instance, execution time or number of intermediate results. These, in turn, depend on estimates of the cardinalities of subqueries.
Ref.~\cite{han2021cardinality} reviews
cardinality estimation techniques and their impact on JO optimisation.
Some approaches apply machine learning for cardinality or cost estimation~\cite{kim2022,hasan2014,akdere2012learning}, to improve the DP optimiser, or to directly determine the JO~\cite{marcus18, krishnan18, xiang20, chen22, trummer21}.

Using quantum approaches to address database problems is a relatively new
field of research, even with early work by Trummer and Koch on solving multi-query
optimisation with quantum annealers only going back to 2016~\cite{trummer16}.
A recent review~\cite{winker:23:sigmodtutorial} summarises existing 
work and classifies potential use-cases. For instance, transaction
scheduling~\cite{groppe21, bittner20a, bittner20b, calikyilmaz23}
schema matching~\cite{Fritsch2023} or tuning index 
configurations~\cite{Gruenwald23} have been addressed 
using quantum methods. 

The join order problem has been cast as an optimisation problem in quadratic unconstrained binary optimisation (QUBO)
form by Schönberger~\etal based on known transformations to
mixed-integer linear programming~~\cite{schoenberger23}, and using a
direct encoding that has also been evaluated on quantum-inspired
hardware~\cite{schoenberger24}. These two solutions for the JO problem are restricted
to left-deep join trees; alternative formulations that allow for handling
general \emph{bushy} join trees were given by Nayak~\etal~\cite{nayak23} and 
Schönberger~\etal~\cite{schoenberger23:qdsm} (we discuss differences in
their scalability in~\autoref{sec:discussion}). Finally, Ref.~\cite{winker23}
introduces an RL inspired approach for the JO problem
using VQCs. It uses rewards to measure the quality of different join orders, but creates a join 
order in a single step and not over multiple interactions with an environment. 

\section{Preliminaries}
\label{sec:preliminaries}

This section introduces the three main concepts relevant to this work, namely the JO problem (\autoref{sec:bg-jo}), and classical (\autoref{sec:bg-rl}) and quantum (\autoref{sec:bg-qrl}) RL.

\subsection{Background on the Join Order Problem}
\label{sec:bg-jo}

The JO problem constitutes of three basic elements:

\subsubsection{Query}
A query formulated in the structured query language (SQL) (see left of \autoref{fig:overview} for an example), can be represented as an expression of relational algebra to be optimised before execution~\cite{Smith1975}.
In this work, we focus on the important problem of JO optimisation with consideration of selection (\ie, filter) operations while the query, in general, may also consist of other operations.
Here, a query $Q$ can be characterised by a join graph and predicates, which can be further decomposed into join predicates and selection predicates.
A join graph for a query is defined by relations that represent the vertices of the graph and filter on which two relations can be joined. These are called join predicates (\eg, \texttt{a1.a=D.a} in \autoref{fig:overview}) and correspond to the edges of the join graph.
The join graph is given by a symmetric adjacency matrix $G \in \mathbb{F}_2^{r\times r}$, where $r$ is the number of relations. If there is a join predicate in $Q$ connecting the relations $r_i$ and $r_j$, the entry $g_{i, j}$ in $G$ is $1$.
Selection predicates are additional filters that act on one relation (\eg, \texttt{D.c > 5} in \autoref{fig:overview}), and can be formalised as described in \autoref{sec:state-rep} or \autoref{sec:mod-reduced}.

\subsubsection{Join Tree}
In contrast to a query graph, which serves as the input for the JO problem, a join tree embodies a solution.
Its leaf nodes represent the base relations to be joined, while its intermediate nodes denote join operations.
Each join node, requiring two operands, has two predecessors: either a) a base relation or b) another join tree node, which itself will be further joined.
The result of a join serves as an operand for another join, indicated by an outgoing edge connecting to its successor.
The only exception is the final join, which does not serve as an operand for any subsequent join.
In this study, we refer to intermediate join trees as "sub-trees".
The top of \autoref{fig:sequence} illustrates the sequence for constructing a complete join tree.

While these requirements apply universally to join trees, certain JO methods impose additional constraints on their structure to enhance efficiency by reducing the search space.
Particularly, some methods exclusively consider \emph{left-deep} join trees, necessitating at least one base relation as an operand for each join.
Consequently, directly joining two pairs of relations is precluded, as it necessitates a join operation on the results of two preceding joins.
Valid left-deep join orders must therefore represent a permutation of relations.
This restriction to left-deep trees was employed in two quantum approaches for JO~\cite{schoenberger23, schoenberger24}.
Nonetheless, the detrimental impact of this constraint on solution quality can be significant, as demonstrated, for instance, by the empirical analysis conducted by Neumann and Radke~\cite{neumann18}.
Hence, our QML approaches consider general or \emph{bushy} join trees, devoid of further structural constraints.
The divergence in scalability between existing quantum-based left-deep and bushy variants is described in~\autoref{sec:scalability}.

\subsubsection{Cost Functions}
Finally, a cost function evaluates the join tree, by assigning it a cost value.
The literature proposes various definitions of cost functions~\cite{kurella18}; some are straightforward yet less precise, while others are more intricate, taking into account multiple factors and closely reflecting real costs (\ie, query execution time including I/O costs).
To evaluate the selected join order, we use the established cost function $C_{\text{out}}$~\cite{cluet95}, which considers the cardinalities (\ie, the number of tuples in a query result set) as an approximation of query complexity:
\begin{equation}
  C_{\text{out}}(T) =
    |T| + C_{\text{out}}(T_1) + C_{\text{out}}(T_2),
\end{equation}
where $n$ is the maximum number of joins in the query, a join tree is defined as $T = T_1 \Join T_2$, and $|T|$ represents the true cardinality of $T$ ($C_{\text{out}}(T) = 0$ if $T \in \{r_1, r_2, \dots\}$ is a leaf).

\subsection{Background on Reinforcement Learning}
\label{sec:bg-rl}

The setup in RL is typically described by the notion of a \textit{Markov decision process} (MDP)~\cite{sutton18}, where an \emph{agent} interacts with an \emph{environment} at discrete time steps $t$.
In each time step, the current configuration of the agent in the environment is summarised by the \textit{state} $S_t \in \mathcal{S}$, where $\mathcal{S}$ is the set of all possible states.
Based on this information, the agent selects an \textit{action} $A_t$ from a set of possible actions $\mathcal{A}$ according to a \textit{policy} $\pi(s,a) = \mathbb{P}[A_t = a\mid S_t = s]$, which gives the probability $\mathbb{P}$ of taking action $a$ in state $s$.
Executing the selected action causes the environment to transition to a next state $S_{t+1} \in \mathcal{S}$.
Simultaneously, the agent receives a scalar \textit{reward} $R_{t+1} \in \mathcal{R}$ that quantifies the contribution of the selected action towards solving the task, with $\mathcal{R} \subset \mathbb{R}$ being the set of all rewards.
$S_{t+1}$ and $R_{t+1}$ are determined by the environment's dynamics $p: \mathcal{S} \times \mathcal{R} \times \mathcal{S} \times \mathcal{A}$, which characterises the probability distribution of a transition $(S_t, A_t, R_{t+1}, S_{t+1})$.

The agent's goal is to maximise the return~\cite{sutton18} $G_t = \sum_{t'=t}^T \gamma^{t'-t} R_{t'+1}$, that is, the discounted sum of rewards, until a terminal timestep $T$ is reached, where the discount factor $\gamma \in (0,1]$ controls how much the agent favours immediate over future rewards.
The period between the initial time step and $T$ is often referred to as an \emph{episode}.

To find a good policy that maximises the return, various RL methods exist~\cite{sutton18}.
As our baseline~\cite{marcus18}, in this work we focus on \emph{Proximal Policy Optimization} (PPO) from the class of policy gradient methods~\cite{schulman17}.
The goal of policy gradient methods is to directly learn the parameterised policy $\pi_{\boldsymbol{\theta}}: \mathcal{S} \times \mathcal{A} \rightarrow [0,1]$, where $\boldsymbol{\theta}$ denote trainable parameters of a function approximator, such as a neural network (NN), or a variational quantum circuit (VQC).
In PPO, the parameters $\boldsymbol{\theta}$ can be optimised using a gradient ascent method, maximising the following objective, consisting of three parts:
\begin{equation}
  \label{eq:objective}
  L_t^{\text{clip}+\text{VF}+\text{S}}(\boldsymbol{\theta}) = \\
  \mathbb{E}_t\left[L_t^{\text{clip}}(\boldsymbol{\theta}) -c_1 L_t^{\text{VF}}(\boldsymbol{\theta}) + c_2 S(\pi_{\boldsymbol{\theta}})\right].
\end{equation}
The PPO algorithm alternates between sampling and optimisation stages. Therefore, $\mathbb{E}_t$ indicates the average over a finite batch of samples, which is gathered prior to each optimisation stage.
$c_1$ and $c_2 \in \mathbb{R}^+$ are hyperparameters.
The clip-objective $L^{\text{clip}}(\boldsymbol{\theta})$, is defined as $r_t(\boldsymbol{\theta})\mathbb{A}_t$, where the ratio $r_t(\boldsymbol{\theta}) = \frac{\pi_{\boldsymbol{\theta}}(a_t,s_t)}{\pi_{\boldsymbol{\theta}_{\text{old}}}(a_t,s_t)}$ is clipped in $1\pm \epsilon$ with $\boldsymbol{\theta}_{\text{old}}$ being the parameters before the update, $\epsilon \in \mathbb{R}$ a hyperparameter and $\mathbb{A}_t$ an advantage estimation of the current policy.
The advantage estimation $\mathbb{A}_t$ itself can be learned by a function approximator based on the value function in an MDP $V(s) = \mathbb{E}_t\left[G_t\mid S_t = s\right]$, that is an estimation of the return, and optimised through the objective $L_t^{\text{VF}}$, which is a squared-error loss function of estimated values from the function approximator and target values, collected in the sampling stage.
The third part in \autoref{eq:objective} $S[\pi_{\boldsymbol{\theta}}]$ denotes the entropy of $\pi_{\boldsymbol{\theta}}$, which is added to ensure sufficient exploration.
For a detailed discussion on PPO, we point readers to Ref.~\cite{schulman17}.

We refer to the policy function approximator that mainly contributes to $L^{\text{clip}}(\boldsymbol{\theta})$ as the \emph{actor}, as it represents the policy that \enquote{acts} in the environment and to the advantage estimator, which is optimised through $L^{\text{VF}}(\boldsymbol{\theta})$ as the \emph{critic}, which evaluates a current policy.
We investigates classical and quantum versions of actor/critic in \autoref{sec:exp-qrl}.

\subsection{Background on Quantum Machine Learning}
\label{sec:bg-qrl}
As a variational quantum circuit (VQC) is proven to be a universal function approximator~\cite{perez20}, similar to a classical NN~\cite{hornik89}, it can be employed as a set-in for NNs in a variety of settings (\eg, \cite{farhi14, mitarai18}), including PPO.
A VQC's structure often follows the data processing flow of a classical NN and comprises three fundamental components:
In the first part, a quantum state is prepared to represent the classical input data~$\boldsymbol{x}$ through applying a unitary gate $\hat{U}_{\text{enc}}(\boldsymbol{x})$ to the initial quantum state, which by convention is $\otimes_n \ket{0}$ for a configuration with $n$ qubits~\cite{nielsen10}.
In the second so-called \emph{variational} part, the quantum state is then transformed by applying a parameterised unitary $\hat{U}_{\text{var}}(\boldsymbol{\theta})$.
An exemplary gate sequence for the encoding and the variational part is depicted in \autoref{fig:vqc_detail}.
Finally, classical information~$\langle \hat{O} \rangle$ is obtained from the quantum circuit by measuring the state.
The notation $\langle \hat{O} \rangle$ refers to the \emph{expectation value} of an observable $\hat{O}$.

The parameters of the VQC are optimised using classical approaches such as
gradient ascent to maximise an objective function, where the gradient of a parameter with respect to the measurement can be calculated using the \emph{parameter-shift rule}~\cite{mitarai18, schuld18}.
As algorithms involving VQCs perform calculations on both, the quantum processing unit (QPU) and CPU, they are called 
\emph{hybrid} approaches.

\subsubsection{Data Encoding}
\label{subsec:enc}
The encoding unitary $\hat{U}_{\text{enc}}(\boldsymbol{x})$ depends on the encoding strategy; Weigold~\etal~\cite{weigold20, weigold21} survey common strategies.
Among these, we focus on \emph{angle encoding}, which uses a Pauli-rotation gate to encode one real value into one qubit.
The corresponding unitary can comprise one (\eg, \cite{skolik22}) or multiple (\eg, \cite{chen20}) parameterised rotation gates per qubit.
Given that the gates are periodic, each input element must be scaled to an interval smaller than $2\pi$.

Even if payload data are not required to encode JO problems, a
simple angle encoding scheme for JO exceeds the capability of
NISQ devices for even small instances.
We therefore employ \emph{incremental data uploading}~\cite{periyasamy22} to spread the encoding gates for the input elements throughout the quantum circuit with parameterised unitaries in between them, which increases \emph{circuit depth} (\ie, the longest gate sequence), but decreases qubit count.
As there is no limit on the maximum number of repetitions of input elements, encoding unitaries can be re-introduced multiple times into the VQC.
This approach, known as \emph{data re-uploading} (DRU)~\cite{perez20}, is suggested to increase the expressivity of a VQC~\cite{schuld21}, which in turn determines the class of functions a VQC can approximate.
In \autoref{sec:exp-qrl} we empirically evaluate and compare the combination of incremental data uploading and DRU.

\subsubsection{Data Decoding}
\label{subsec:dec}
Several techniques are known to map \enquote{outputs} of a 
VQC (\ie, the expectation value of multiple measurements) to a set 
of output values that is smaller than or equal to the number of 
qubits~\cite{jerbi21, meyer23}.
Few existing approaches~\cite{lockwood21} decode 
quantum states to larger output spaces.
As described in \autoref{sec:meth-rj}, action and output space are typically larger than the number of qubits for JO.
We therefore determine the expectation value for each qubit individually using $\hat{Z}$ observables and feed the outcomes into one classical NN layer with the correct output size for the actor.
For the critic model, which only requires one output component to estimate the advantage, circuit outcome is determined by observable $\otimes_n \hat{Z}$.
Since the expectation value of 
$\hat{Z}$ lies in $[-1,1]$ the critic model outcome is scaled using an additional trainable classical parameter and bias.

\section{Methodology}
\label{sec:methodology}
To understand how RL can be utilised for the JO problem on QCs, 
we commence with discussing the differences between building the join order step-wise or returning the full join order within one single step using a machine learning (ML) model. We also introduce a single-step approach based on QML.
Subsequently, we outline our classical baseline \rj and the adjustments required for quantum RL.

\subsection{Single-Step versus Multi-Step Join Ordering}
\label{sec:meth-single-versus-multi-step}

\definecolor{blond}{rgb}{0.98, 0.94, 0.75}
\definecolor{RLTerms}{rgb}{0.0, 0.48, 0.65}

\begin{figure}[htb]
  \centering
  \input{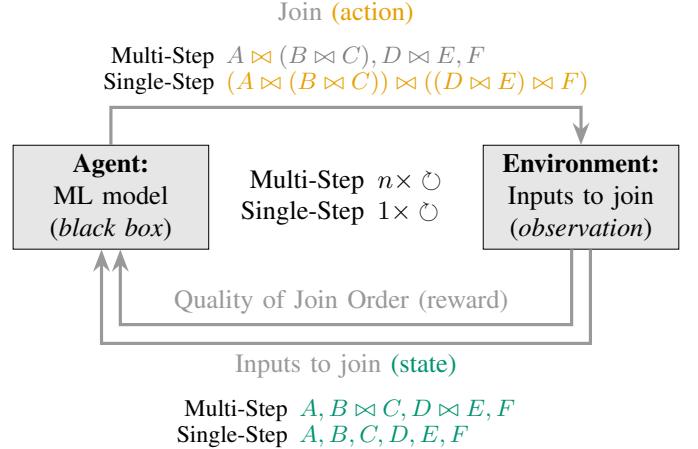}

\tikzset{block/.style={rectangle, draw, very thin, minimum width=2.6cm,
                       align=center, fill=bggray}}
\def\postLabSep{0.75em}

\begin{tikzpicture}
  \renewcommand{\tabcolsep}{0.25em}

  \node[block, anchor=west] (a) at (0, 0) {
     \textbf{\color{black}Agent:}\\
     ML model\\
     (\emph{black box})
  };

  \node[block, anchor=east] (e) at (\linewidth, 0) {
     \textbf{\textcolor{black}{Environment:}}\\
     Inputs to join\\
     (\emph{observation})
  };

  \draw[infoStyle] ($(e.south) + (0.125cm, 0)$) -- 
                   ($(e.south) + (0.125cm, -1.25cm)$) -- 
                   ($(a.south) + (-0.125cm, -1.25cm)$) 
                   node[midway, below, align=center] {
                      Inputs to join \textcolor{lfd4}{(state)}\\[\postLabSep]
                      {\small\begin{tabular}{rl}
                         \textcolor{black}{Multi-Step} & $\mathcolor{lfd4}{A, B\bowtie C, 
                                                                 D\bowtie E, F}$\\
                         \textcolor{black}{Single-Step} & $\mathcolor{lfd4}{A, B, C, D, E, F}$
                      \end{tabular}}
                    } -- ($(a.south) + (-0.125cm, 0)$);

  \draw[infoStyle] ($(e.south) + (-0.125cm, 0)$) -- 
                   ($(e.south) + (-0.125cm, -1cm)$) -- 
                   ($(a.south) + (0.125cm, -1cm)$) 
                   node[midway,above,align=center]{
                      Quality of Join Order (reward)
                   } -- ($(a.south) + (0.125cm, 0)$);

  \draw[infoStyle] (a) -- ($(a.north) + (0, 0.5cm)$) -- ($(e.north) + (0, 0.5cm)$) 
                   node[midway,above,align=center] {
                   Join \textcolor{lfd2}{(action)}\\[\postLabSep]
                      {\small\begin{tabular}{rl}
                         \textcolor{black}{Multi-Step} & 
                                               \(\color{gray}A\mathcolor{lfd2} 
                                               {\bowtie}\color{gray}(B\bowtie C), 
                                               D\bowtie E, F\)\\
                         \textcolor{black}{Single-Step} & 
                         \(\mathcolor{lfd2}{(A\bowtie 
                                                (B\bowtie C))\bowtie((D\bowtie E)\bowtie F)}\)
                      \end{tabular}}
                   } -- (e);

  \node[align=center, anchor=center] (middle) at (\linewidth/2, 0) { 
       \begin{tabular}{rl}
         Multi-Step  & $n\times\circlearrowright$\\
         Single-Step & $1\times\circlearrowright$
       \end{tabular}
  };
\end{tikzpicture}
  \vspace*{-2em}
  \caption{Single-step versus multi-step approach presented in an RL fashion. Here, $A$ to $F$ are the relations to join. We neglect selection predicates.}\label{fig:SingleVersusMulti}
\end{figure}

A join tree can be created by an ML model in multiple steps or in a single step. \autoref{fig:SingleVersusMulti} summarises the differences:
The state of the environment contains
(1) already determined subjoins, (2) relations to be joined, and (3) selection predicates.
An ML model acts as an agent in the RL context. It predicts and emits the next best subjoin (\ie, the action) connecting two of the subjoins and relations of the previous state.
Thereby, the environment of already determined subjoins and to be joined relations is updated.
By determining the quality (\ie, the reward) of the intermediate join order, the model can be trained to better predict the next best subjoin.
In the \emph{multi-step approach}, a single join is added to the join tree in each step until all input relations are joined (\ie, a terminal state is reached).
In the \emph{single-step approach}~\cite{winker23}, the model directly generates
a complete join order without intermediate steps (\ie, a terminal state is
reached after one step).

\subsection{Single-Step QML}
\label{sec:meth-single-step}

In the single-step QML approach \cite{winker23}, all join orders are enumerated, and each join tree is associated with a quantum state. The join order associated with the most commonly measured quantum state is taken as join tree. The quality of join orders can vary greatly and multiple join orders can have equal or nearly equal good quality. For instance, the second-best join order might only be slightly worse, while another might have a large difference in quality. Thus, it is not a good approach to make a binary choice between right or wrong for a join order. Instead, each join order is assigned a reward depending on its quality. We use the VQC to predict these rewards and choose the join order with the highest reward.
\subsection{Multi-Step QRL}

\def\faca{0.4}
\def\facb{0.4}
\def\facc{0.05}
\def\boxSkip{0.75em}
  \newbox\stepa\newbox\stepb
  \setbox\stepa=\hbox{\begin{minipage}{\faca\linewidth}\begin{center}
\begin{tikzpicture}
\setlength{\arraycolsep}{0.5pt}
\node[anchor=south west] (mtx) at (0,0) { \(
\footnotesize\begin{bNiceMatrix}[first-row,code-for-first-row=\footnotesize,
                    first-col,code-for-first-col=\tiny]
              & a_{1} & a_{2} & B & C  & D \\
        a_{1} & 0 & 0 & 0 & 0 & 1 \\
        a_{2} & 0 & 0 & 0 & 0 & 1\\
        B     & 0 & 0 & 0 & 0 & 0\\
        C     & 0 & 0 & 0 & 0 & 0\\
        D     & 1 & 1 & 0 & 0 & 0
  \end{bNiceMatrix}\)};
  \node[rotate=90,anchor=south] at ($(mtx.south west)!2.5/6!(mtx.north west)$) { \emph{Join Graph} };
\end{tikzpicture}\vspace*{\boxSkip}

     \emph{Selection Predicates}\\
     (0, 0, 0,\ldots, 0, 1, 0)\vspace*{\boxSkip}
     
\begin{tikzpicture}
\setlength{\arraycolsep}{0.5pt}
\node[anchor=south west] (mtx) at (0,0) { \(
       \footnotesize\begin{bNiceMatrix}[
          first-row,code-for-first-row=\tiny,
          first-col,code-for-first-col=\tiny,
        ]
        & a_{1} & a_{2} & B & C  & D \\
        a_{1} & 1 & 0 & 0 & 0 & 0 \\
        a_{2} & 0 & 1 & 0 & 0 & 0\\
        B     & 0 & 0 & 0 & 0 & 0\\
        C     & 0 & 0 & 0 & 0 & 0\\
        D     & 0 & 0 & 0 & 0 & 1
        \end{bNiceMatrix}\)};
  \node[rotate=90,anchor=south] at ($(mtx.south west)!2.5/6!(mtx.north west)$) { \emph{Tree Structure} };
\end{tikzpicture}

\end{center}\end{minipage}}
  \setbox\stepb=\hbox{\begin{minipage}{\facb\linewidth} \begin{center}
\begin{tikzpicture}
\setlength{\arraycolsep}{0.5pt}
\node[anchor=south west] (mtx) at (0,0) { \(
       \footnotesize\begin{bNiceMatrix}[
          first-row,code-for-first-row=\tiny,
          first-col,code-for-first-col=\tiny,
        ]
              & a_{1} & a_{2} & D \\
        a_{1} & 0 & 0 & 1 \\
        a_{2} & 0 & 0 & 1 \\
        D     & 1 & 1 & 0
        \end{bNiceMatrix}\)};
  \node[rotate=90,anchor=south,align=center] at ($(mtx.south west)!2.5/6!(mtx.north west)$) { \emph{Join}\\\emph{Structure} };
  \end{tikzpicture}\vspace*{\boxSkip}

      \emph{Selection Predicates}\\
       Indices: \((0,0,3)\)\\
       Selectivities: \((1,1,0.75)\)\vspace*{\boxSkip}

\begin{tikzpicture}
\setlength{\arraycolsep}{0.5pt}
\node[anchor=south west] (mtx) at (0,0) { \(
       \footnotesize\begin{bNiceMatrix}[
          first-row,code-for-first-row=\tiny,
          first-col,code-for-first-col=\tiny,
        ]
        & a_{1} & a_{2} & D \\
        a_{1} & 1 & 0 & 0 \\
        a_{2} & 0 & 1 & 0 \\
        D     & 0 & 0 & 1
        \end{bNiceMatrix}\)};
  \node[rotate=90,anchor=south,align=center] at ($(mtx.south west)!2.5/6!(mtx.north west)$) { \emph{Tree}\\\emph{Structure} };
\end{tikzpicture}

    \end{center} \end{minipage}}
\begin{figure*}[htb]
  \hspace*{-6.5mm}\begin{minipage}{\linewidth}
  \input{tikz-figs/dimensions.tex}

\begin{tikzpicture}[remember picture]
  \coordinate (tl) at (0, \HEIGHT);
  \coordinate (tr) at (\WIDTH, \HEIGHT);
  \coordinate (bl) at (0, 0);
  \coordinate (br) at (\WIDTH, 0);


  \draw (0,\HEIGHT) node (qbox) [inner sep=0pt, rotate=90, anchor=north east] {
    \begin{minipage}{0.8\HEIGHT}
      \begin{querybox}{Query}
        \begin{lstlisting}[style=query]
SELECT * FROM A as a1, A as a2, D
WHERE a1.a=D.a AND a2.b=D.b
               AND D.c > 5;
        \end{lstlisting}
      \end{querybox}
    \end{minipage} 
  };

\coordinate (tstart) at (qbox.south east);
\calcLength(tstart,tr){tspace}
\newlength{\dist}\setlength{\dist}{\tspace pt}
\newlength{\dbwidth}
\pgfmathsetmacro{\tmp}{(\dist-(\facc\linewidth+\wd\stepa+\wd\stepb))/3-2mm}
\setlength{\dist}{\tmp pt}
\pgfmathsetmacro{\tmp}{\dist+0.5*(\faca\linewidth+\facb\linewidth)}
\setlength{\dbwidth}{\tmp pt}

  \node[anchor=north west, fill=bggray, minimum width=\dbwidth] (db)
        at ($(tstart)+(\dist, 0)$) { 
        \begin{tabular}{ll}
          Tables &\(\{A, B, C, D\}\)\\
           Aliases &\(\{a_{1}, a_{2}, B, C, D\}\)\\
        Attributes &\(a_{1}.a, a_{1}.b,\ldots, D.c, D.d\)
        \end{tabular}
        };
  
  \node[very thin, anchor=north west, fill=bggray, inner sep=0pt] (data1) at 
       (db.south west) { \usebox{\stepa} };
  \node[very thin, anchor=north east, fill=bggray, inner sep=0pt] (data2) at
       (db.south east) { \usebox{\stepb} };
%
  \newcommand{\ca}[1]{\mathcolor{lfd1}{#1}}
  \newcommand{\cb}[1]{\mathcolor{lfd2}{#1}}
  \newcommand{\cc}[1]{\mathcolor{lfd4}{#1}}
  \node[anchor=north west, inner sep=0pt] (data3) at 
                                ($(db.north east) + (\dist, 0)$) { 
       \begin{minipage}{\facc\WIDTH}
         \begin{displaymath}
           \footnotesize\begin{bNiceMatrix}
           \ca0\\\ca0\\\ca3\\\ca{1.0}\\\ca{1.0}\\\ca{0.75}\\\ca1\\\ca0\\\cb0\\\cb0\\\cb1\\\cb0\\\cb0\\\cb0\\\cb1\\\cb0\\\cc0\\\cc1\\\cc0\\\cc0\\\cc1\\\cc1\\\cc1\\\cc0
           \end{bNiceMatrix}
         \end{displaymath}
       \end{minipage}
  };

  \coordinate (m1) at ($(data3.south east)!0.85!(data3.north east)$);
  \coordinate (m2) at ($(data3.south east)!0.50!(data3.north east)$);
  \coordinate (m3) at ($(data3.south east)!0.15!(data3.north east)$);

    \path[infoStyle] ([xshift=\lineDist]qbox.south) -- +(\dist-\lineDist, 0) 
                         node[above, midway]{ Baseline } 
                         node[below, midway]{ Encoding };

  \coordinate (sarrow) at ($(data1.north east)!(qbox.south)!(data1.south east)$);      
  \path[infoStyle] ([xshift=\lineDist]sarrow) -- +(\dist-\lineDist, 0) 
                         node[above, midway]{ Feature } 
                         node[below, midway]{ Reduction };                         

  \coordinate (sarrow) at ($(data2.north east)!(qbox.south)!(data2.south east)$);      
  \path[infoStyle] ([xshift=\lineDist]sarrow) -- +(\dist-\lineDist, 0) 
                         node[above, midway]{ Flattening \& } 
                         node[below, midway]{ Concatenation };

\yquantdefinebox{dots}[inner sep=0pt]{$\dots$}
\hspace*{\circShift}\input{tikz-figs/circuit.tex}
\end{tikzpicture}

\begin{tikzpicture}[remember picture, overlay]

\coordinate (start_var) at ($(sc1-var1.south) - (0,\varCtrlDist)$);
\coordinate (tmp) at ($(pp.north east)!(meas1.east)!(pp.north east)$);
\coordinate (tr_var) at ($(tmp) + (\varCtrlDist, 0)$);
\coordinate (corner_var) at (start_var -| tr_var);
\path[varStyle] (start_var) -| (tr_var);

\foreach \n in {1,4} { 
    \path[varStyle] (meas\n.east) -- ($(pp.north west)!(meas\n.east)!(pp.south west)$)
                    ($(pp.north east)!(meas\n.east)!(pp.north east)$) -- 
                   +(\varCtrlDist, 0);
}

\foreach \n in {1,2,3} { 
  \foreach \m in {1,2,3} { 
    \path[varStyle, -Stealth] ($(sc\n-var\m.south) - (0, \varCtrlDist)$) -- 
                              (sc\n-var\m.south);
  }
}

\coordinate (l1) at ($(m1) + (\varCtrlDist, 0)$);
\coordinate (l2) at ($(m3) + (\varCtrlDist, 0)$);
\coordinate (prj) at ($(sc1-enc1.north) + (0, \varCtrlDist)$);
\coordinate (corner_enc) at ($(corner_var)!(prj)!(tr_var)$);
\coordinate (start_enc) at (m1 -| corner_enc);


\foreach \n in {1,2,3} { 
    \path[thick, dataStyle] ([xshift=-\circShift]m\n) -- 
                           ($(start_enc)!(m\n)!(corner_enc)$);
}
\path[dataStyle] (start_enc) -- (corner_enc);

\foreach \n in {1,2,3} { 
  \foreach \m in {1,2,3} { 
    \path[dataStyle, -Stealth] (corner_enc) -| (sc\m-enc\n.north);
  }
}

\node[anchor=south east] at (corner_var.north west) { \textbf{\(\circlearrowright\)} };
\end{tikzpicture}
  \end{minipage}
  
  \caption{Interplay between data encoding (top) and variational quantum circuit
  (bottom) processing in our approach. Starting from the query and the baseline
  encoding of Ref.~\cite{marcus18}, we prune unnecessary features 
  and flatten the core input data into a vector that is statically
  fed into the encoding quantum gates \(\hat{U}_{\text{enc}}\). The variational quantum
  circuit (using a configurable number of qubits) is initialised with qubits in 
  state \(\ket{0}\), and iteratively executes block of intermingled encoding
  and variational (\(\hat{U}_{\text{var}}\)) gates; following a measurement, a 
  classical optimisation procedure delivers new parameter estimates for the variational 
  gates, and the updated circuit is iteratively re-executed.
  Following established conventions, solid lines indicate quantum information, 
  double lines concern classical information (measurement results that may change in each run of 
  the quantum circuit), and dashed lines represent parameters that are statically fed into the 
  quantum circuit (remaining constant across circuit runs). Grey, thick lines symbolise 
  logical flow.}\label{fig:overview}
\end{figure*}

\autoref{fig:overview} visualises our QRL multi-step approach.
By using a state representation based on Ref.~\cite{marcus18}, a VQC can choose the next join in an iterative process until a complete join order is built.
The classical baseline as well as the modifications required for the application of QRL are described below.

\subsubsection{Classical Baseline---\rj}
\label{sec:meth-rj}

For our multi-step approach, we utilised the method described in Ref.~\cite{marcus18}.
Although the literature proposes various RL methods for the JO problem (cf. \autoref{sec:rel-work}), we opted for \rj as a foundation because of its compact input feature space.
Other approaches, such as \emph{RTOS}~\cite{xiang20} or \emph{JOGGER}~\cite{chen22}, utilise sophisticated classical machine learning techniques to represent states of queries and databases, which lack a direct equivalent in the domain of quantum computing.
Investigating novel methods that apply these advanced classical machine learning techniques to a quantum domain is beyond the scope of this study.
Instead, our QRL approach should evaluate the capabilities of existing QML methods on small input spaces of the JO problem to establish a lower bound for the potential of using QRL, or QC in general.
Additionally, due to the limited number of qubits on current NISQ devices and each quantum circuit gate being a potential source for noise and imperfections, it is beneficial to reduce the classical data encoded into the quantum gates to a minimum.
As outlined below, \rj employs a total of $a + 2r^2$ input features, where $r$ denotes the number of tables and $a$ represents the total number of attributes in the database with $a > r$. As we show in \autoref{sec:mod-reduced} we are able to reduce the input space even further.
In contrast, for example \emph{DQ}~\cite{krishnan18} necessitates roughly $r \times (a + 1)$ features, resulting in a larger input space considering that the number of attributes typically outweighs the number of relations in the database.

\paragraph{MDP}
The MDP's state for the JO problem is represented by a query $Q$ and a set of relations or (sub-)join-trees $\mathcal{F}$.
The PPO agent sequentially combines two sub-trees $T_k, T_l \in \mathcal{F}$, which corresponds to an action, until a complete join order is build.
Building the join order for one query, represents an episode.
The agent aims for a join order that achieves minimum costs respectively a maximum reward.

\begin{figure*}[htb]
 \input{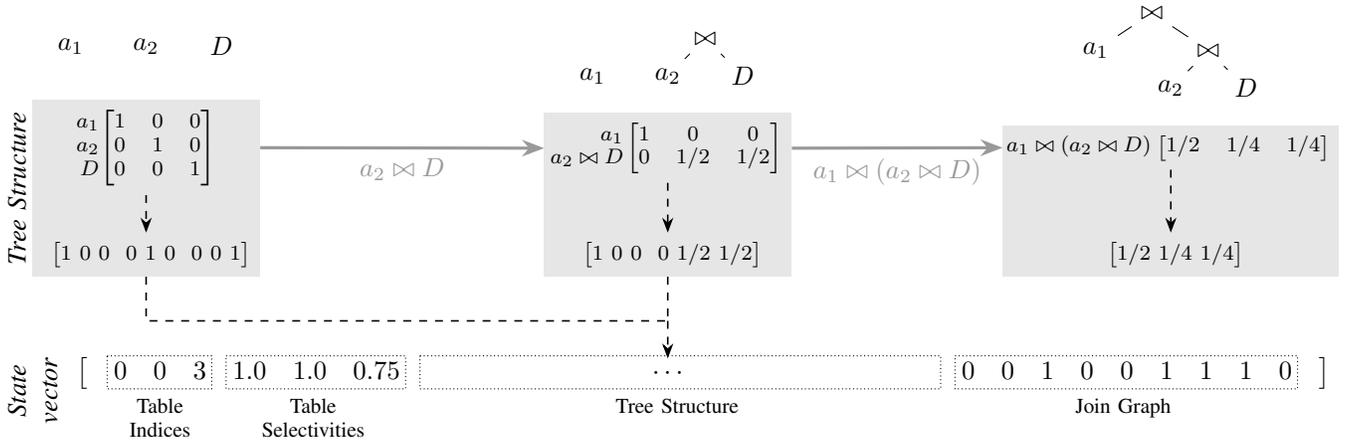}

\newcommand{\enclose}[3]{%
    \draw[densely dotted, thin] ([xshift=0.1em,yshift=-0.1em]#1.north west) -| 
                        ([xshift=-0.15em]#2.south east) -| cycle;
    \node[anchor=north,align=center,font=\footnotesize] at 
                        ($(#1.south)!0.5!(#2.south)$) { #3 };
}

\tikzset{joNode/.style={}}
\tikzset{joEdge/.style={draw}}
                          
\begin{tikzpicture}[remember picture]
\pgfdeclarelayer{background}\pgfsetlayers{background,main}
\setcounter{MaxMatrixCols}{20}
      \node[anchor=south west] (state) at (0,0) {
         \(\begin{bmatrix} &
            \subnode{istart}{0} & 0 & \subnode{iend}{3} & 
            \subnode{sstart}{1.0} & 1.0 & \subnode{send}{0.75} &
            \subnode{tstart}{\hspace{3cm}\vphantom{0}}  \subnode{tmid}{\cdots} &
            \subnode{tend}{\hspace{3cm}\vphantom{0}}\strut & \subnode{jstart}{0} &
            0 & 1 & 0 & 0 & 1 & 1 & 1 & \subnode{jend}{0} & \end{bmatrix}\)
      };

      \enclose{istart}{iend}{Table\\Indices}
      \enclose{sstart}{send}{Table\\Selectivities}
      \enclose{tstart}{tend}{Tree Structure}
      \enclose{jstart}{jend}{Join Graph}

      \node[rotate=90, align=center, font=\itshape, anchor=south east] (svlabel) at (state.north west) { State\\vector };

    \node[anchor=west] (j1) at ([yshift=3cm]state.west) { \(
    \footnotesize\begin{bNiceMatrix}[first-col,code-for-first-col=\tiny]
        a_{1} & 1 & 0 & 0 \\
        a_{2} & 0 & 1 & 0 \\
        D     & 0 & 0 & 1
                 \end{bNiceMatrix}\)
    };
    
    \node[anchor=center] (j2) at ([yshift=3cm]tmid) { \(
    \footnotesize\begin{bNiceMatrix}[first-col,code-for-first-col=\tiny]
        a_{1}           & 1 & 0 & 0 \\
        a_{2} \Join D   & 0 & 1/2 & 1/2
                 \end{bNiceMatrix}\)
    };

   \node[anchor=east] (j3) at ([yshift=3cm]state.east) { \(
    \footnotesize\begin{bNiceMatrix}[first-col,code-for-first-col=\tiny]
        a_{1} \Join (a_{2} \Join D) & 1/2 & 1/4 & 1/4
                 \end{bNiceMatrix}\)
    };

  \node[joNode, anchor=south] (j1_a2) at ([yshift=\varCtrlDist]j1.north) { \(a_{2}\) };
  \node[joNode] (j1_a1) at ([xshift=-\joNodeDist]j1_a2) { \(a_{1}\) };
  \node[joNode] (j1_D) at  ([xshift=\joNodeDist]j1_a2) { \(D\) };

  \node[joNode] (j2_a2) at ([yshift=\varCtrlDist]j2.north) { \(a_{2}\) };
  \node[joNode] (j2_a1) at ([xshift=-\joNodeDist]j2_a2) { \(a_{1}\) };
  \node[joNode] (j2_D) at  ([xshift=\joNodeDist]j2_a2) { \(D\) };  
  \node[joNode] (j2_j1) at  ($(j2_a2)!0.5!(j2_D) + (0, \varCtrlDist)$) { \(\Join\) };  
  \path[draw] (j2_a2) -- (j2_j1) (j2_D) -- (j2_j1);

  \node[joNode] (j3_a2) at ([yshift=\varCtrlDist]j3.north) { \(a_{2}\) };
  \node[joNode] (j3_a1) at ([xshift=-\joNodeDist, yshift=\varCtrlDist]j3_a2) { \(a_{1}\) };
  \node[joNode] (j3_D) at  ([xshift=\joNodeDist]j3_a2) { \(D\) };  
  \node[joNode] (j3_j1) at ($(j3_a2)!0.5!(j3_D) + (0, \varCtrlDist)$) 
                           { \(\Join\) };  
  \node[joNode] (j3_j2) at ($(j3_a1)!0.5!(j3_j1) + (0, \varCtrlDist)$) 
                           { \(\Join\) };                             
  \path[draw] (j3_a2) -- (j3_j1) (j3_D) -- (j3_j1);
  \path[draw] (j3_a1) -- (j3_j2) (j3_j1) -- (j3_j2);

  \node[anchor=north] (j1_vec) at ([yshift=-\varCtrlDist]j1.south) {
    \setlength{\arraycolsep}{1.5pt}
    \footnotesize\(\begin{bmatrix}
        1 & 0 & 0 && 0 & 1 & 0 && 0 & 0 & 1
    \end{bmatrix}\)
  };

  \node[anchor=north] (j2_vec) at ($(j2.north)!(j1_vec.north)!(j2.south)$) {
    \setlength{\arraycolsep}{1.5pt}
    \footnotesize\(\begin{bmatrix}
        1 & 0 & 0 && 0 & 1/2 & 1/2
    \end{bmatrix}\)
  };  

  \node[anchor=north] (j3_vec) at ($(j3.north)!(j1_vec.north)!(j3.south)$) {
    \setlength{\arraycolsep}{1.5pt}
    \footnotesize\(\begin{bmatrix}
        1/2 & 1/4 & 1/4
    \end{bmatrix}\)
  };

  \begin{pgfonlayer}{background}
      \node[fill=bggray, inner sep=0pt,
            fit=(j1) (j1_vec)] (bg1) { };
      \node[fill=bggray, inner sep=0pt,
            fit=(j2) (j2_vec)] (bg2) { };  
      \node[fill=bggray, inner sep=0pt,
            fit=(j3) (j3_vec)] (bg3) { };
  \end{pgfonlayer}

  \coordinate (trgt) at ([yshift=\varCtrlDist]tmid.north);

  \path[dataStyle, -Stealth] (trgt) -- (tmid.north);
  \path[dataStyle, -Stealth] (j3.south) -- (j3_vec.north) ;
  
  \foreach \n in {1,2} {
     \path[dataStyle, -Stealth] (j\n.south) -- (j\n_vec.north) ;
     \path[dataStyle] (j\n_vec.south) |- (trgt) ;
  }
                   
  \path[infoStyle] ($(bg1.south east)!(j1.east)!(bg1.north east)$) -- (j2.west) 
                   node[below, midway]{ \(a_{2}\Join D\) };
  \path[infoStyle] ($(bg2.south east)!(j2.east)!(bg2.north east)$) -- (j3.west) 
                   node[below, midway]{ \(a_{1} \Join (a_{2}\Join D)\) };

  \coordinate(tmp) at ($(j1_vec.south)!0.5!(j1.north)$);
  \node[anchor=north, rotate=90, font=\itshape] at 
                ($(svlabel.north west)!(tmp)!(svlabel.north east)$)
                { Tree Structure } ;

\end{tikzpicture}
 \vspace*{-0.5em}
  \caption{Processing sequence to iteratively determine join orders. Once the query has been parsed and encoded, subsequent invocations of the variational
  quantum circuit as illustrated in~\autoref{fig:overview},
  determine more and more joins, until a complete order has been found.}\label{fig:sequence}
\end{figure*}

\paragraph{State Representation}
\label{sec:state-rep}
Formally, one part of the state representation is the join graph $G$, defined in \autoref{sec:bg-jo}.
Additionally, selection predicates in the query $Q$ are represented by a vector of length $a$, which is the number of attributes in the database.
Selection predicates are one-hot encoded: If a predicate is present in $Q$, the corresponding value in the predicate vector $P$ is one; otherwise zero.
Furthermore, each intermediate sub-tree $T_k \in \mathcal{F}$, that is the tree structure, is encoded as a row vector $\tau_k$ of size $r$.
If a relation $r_i$ is equal to $T_k$ ($r_i$ is a leaf) or is present in $T_k$, then the corresponding value in the row vector $\tau_{k,i}$ is $\frac{1}{h(i, k)}$, where $h(i, k)$ is the height of $r_i$ in $T_k$.
To ensure an evenly sized input space throughout the training process, for each subtree $T_k$ that is successfully joined to another subtree $T_l$, $\tau_{k}$ is set to $\vec{0}$.
There exist $r$ sub-tree row vectors $\mathring{T}$ in total, since at the beginning of each join-process each relation correspond to one sub-tree.
An exemplary sequence of row vectors that is encountered until a full join order is built is depicted in \autoref{fig:sequence}, which 
uses the reduced encoding introduced in \autoref{sec:mod-reduced}.
The complete state for the baseline can be expressed through concatenation, $S_t = G^f \oplus P \oplus (\oplus_{\tau_k \in \mathring{T}} \tau_k)$, where $G^f$ denotes the flattened join graph as a vector and $\oplus$ concatenation with $|S_t| = a + 2r^2$.

\paragraph{Action Representation}
The PPO actor returns a probability distribution over all actions $A_t \in \mathcal{A}$.
The set of actions $\mathcal{A}$ comprises all combinations of two sub-trees $(T_k, T_l) \forall T_k, T_l \in \mathcal{F}, k \neq l$, resulting in an action space of size $r \times (r -1)$.
It encompasses actions with relations that are not present in the query, or lead to a cross join (\ie a join between relations that are not connected by a join predicate). As these typically involve high costs, we apply a mask to the policy by multiplying each value that represents an invalid action with zero to prevent them from being sampled.

\paragraph{Reward Signal}

In previous studies on RL for JO (\eg, Refs.~\cite{krishnan18, marcus18, xiang20}) the reward, as function of cost, is only assigned at the end of each episode when the full join order is built by the RL policy. Intermediate steps receive a zero reward.
This seems counter-productive, given that one property of RL is to determine an action based on a current state and reward signal\footnote{We provide a comparison to a method, which awards zero to intermediate steps in the supplementary material in our \repro}.
Therefore, we propose a multi-step reward signal:
Assuming the cost difference $C_t$ between timesteps $t$ and $t-1$ with costs $c_k$ for subtrees $T_k \in \mathcal{F}_t$ in a state $S_t$ is
\begin{equation}
  C_t = 
  \begin{cases}
    \sum_{T_k \in \mathcal{F}_t} c_k - \sum_{T_l \in \mathcal{F}_{t-1}} c_l & \text{if } t > 0 \\
    0 & \text{if } t = 0
  \end{cases},
\end{equation}
and the cost assigned to the best join order of the full query determined by a DP exhaustive search is $C_{\text{DP}}$, we propose the clipped reward at $t$ as
\begin{equation}
  R_t = \frac{1}{n-1} \left[-\min\left(\frac{C_t}{C_{\text{DP}}}, n-1 \right) + 2\right].
\end{equation}

This requires $n-1$ joins (and actions) to build the join tree for a query with $n$ relations.
Clipping, shifting and normalising the ratio reduces the chances of steeper gradients during training, which is a known cause of suboptimal training~\cite{laud03}.

\subsubsection{Quantum \rj}
For \rj, a VQC can be employed as the actor-, as well as critic-part of PPO, or both.
In both cases, the VQC encodes the state $S_t$. Policy or advantage estimations are obtained using the approach of \autoref{subsec:dec}.

As the number of inputs that a QPU can process is restricted by the hardware capabilities of QPUs, it is advantageous to minimise this
number.
As described in~\autoref{sec:meth-rj}, the state representation of the classical baseline suggests a state space with $a + 2r^2$ features for a database with $r$ relations\footnote{We assume $r$ is the number of different aliases occurring in the dataset, and  $a$ is the number of attributes corresponding to these aliases. One author of Ref.~\cite{marcus18} confirmed that multi-aliases were handled as an additional tables.} and $a$ attributes.
For the JOB, which encompasses 208 attributes across 39 different aliases throughout the JOB query set, there are \numprint{3250} input elements for one state.

\paragraph{Reducing the Input Size}
\label{sec:mod-reduced}
To reduce the observation space, we specify a maximum number of relations $n$ that can be joined.
As for the baseline, we employ a join graph and a tree structure representation, which are defined analogous to the baseline over the $n$ relations present in a given query. This leads to $n^2$ elements in both, the join graph and the sub-tree structure representation.
To specify, which tables are referenced in a query, the tables in the database are enumerated and assigned with an index $I: \mathcal{T} \rightarrow [0, r-1]$, where $\mathcal{T}$ is the set of all tables and $r$ is the number of tables in the database.
The indices $i \in \bigcup_{\mathcal{T}_q \in Q} I(\mathcal{T}_q)$ for a query $Q$ are added to the input components.
To represent the information, which is given through the selection predicates, we obtain the selectivity (\ie, the fraction of tuples present in a result when filtering for the corresponding selection predicates of a specific table) for every table in a query and add these to the input components.

The reduced state representation leads to $2 (n^2 + n)$ input 
elements. For $n=17$ as maximum size in the JOB, 
this results in 612 elements, over 80\% less than
in the baseline.

\begin{figure}[htb]
    \begin{center}
      \scriptsize{\begin{tikzpicture}
  \pgfdeclarelayer{background}\pgfsetlayers{background,main}
  \begin{yquant}
    qubit {} q[4];

    [value=5mm] hspace q[0];
    [value=5mm] hspace q[1];
    [value=5mm] hspace q[2];
    [value=5mm] hspace q[3];
    
    [name=enc0, value=\(\hat{R}_{x}(\theta_{0})\)] box q[0];
    [value=\(\hat{R}_{x}(\theta_{1})\)] box q[1];
    [value=\(\hat{R}_{x}(\theta_{2})\)] box q[2];
    [name=enc1, value=\(\hat{R}_{x}(\theta_{3})\)] box q[3];

    [value=5mm] hspace q[0];
    [value=5mm] hspace q[1];
    [value=5mm] hspace q[2];
    [value=5mm] hspace q[3];

    [name=var0, value=\(\hat{R}_{y}(\gamma_{0})\)] box q[0];
    [value=\(\hat{R}_{y}(\gamma_{1})\)] box q[1];
    [value=\(\hat{R}_{y}(\gamma_{2})\)] box q[2];
    [value=\(\hat{R}_{y}(\gamma_{3})\)] box q[3];

    [value=\(\hat{R}_{z}(\delta_{0})\)] box q[0];
    [value=\(\hat{R}_{z}(\delta_{1})\)] box q[1];
    [value=\(\hat{R}_{z}(\delta_{2})\)] box q[2];
    [name=var1, value=\(\hat{R}_{z}(\delta_{3})\)] box q[3];
    
    [name=ent0] zz (q[0,1]);
    [name=ent1] zz (q[1,2]);
    [name=ent2] zz (q[2,3]);
    [name=ent3] zz (q[3,0]);
    
    [value=5mm] hspace q[0];
    [value=5mm] hspace q[1];
    [value=5mm] hspace q[2];
    [value=5mm] hspace q[3];
  \end{yquant}
  \begin{pgfonlayer}{background}
    \node[draw, dashed, fit=(enc0) (enc1)] (enc) { }; 
    \node[draw, dashed, fit=(var0) (var1) (ent3)] (var) { };
  \end{pgfonlayer}
  \node[anchor=north] (enc_label) at (enc.south) { Encoding \(\hat{U}_{\text{enc}}\)};
  \node[anchor=north] (var_label) at (var.south) { Variational \& Entangling  \(\hat{U}_{\text{var}}\)};
\end{tikzpicture}}
    \end{center}
    \vspace*{-1em}
    \caption{Details of quantum state manipulation: Parametrised 
    rotations around the \(x\) axis (\(\hat{R}_{x}(\theta)\)) 
    encode information. The 
    variational part comprises parametrised rotations around the \(y\) 
    and \(z\) axes, implemented by \(\hat{R}_{y}(\gamma)\) and \(\hat{R}_{z}
    (\delta)\), followed by a cyclic sequence of \CZgate gates that create entanglement.}\label{fig:vqc_detail}
\end{figure}
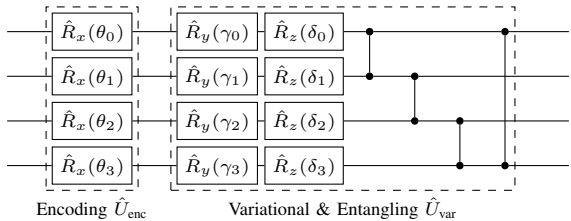

\paragraph{Circuit dimensions}
\label{subsec:dimensions}
One advantage of quantum algorithms involving VQCs is that they allow for a certain degree of controllability of the circuit depth and number of qubits, which is especially desirable for NISQ devices~\cite{preskill18}.
Utilising the incremental data-uploading~\cite{periyasamy22} and the DRU~\cite{perez20} approaches, we can choose the structure of the quantum circuit.
We opted to divide the $2 (n^2 + n)$ input features in $n$ equally sized parts $p_l$.
Each feature $f_i \in p_l$ is then scaled to a range of $[0, \pi]$ and used as s rotation angle for a $\hat{R}_x$ gate acting on qubit $i$ in the \emph{layer} $l$.
The input parts are interleaved with parameterised gates $\hat{R}_y$ and $\hat{R}_z$ that act on each qubit and introduce trainable parameters, and a circular sequence of \CZgate gates between two adjacent qubits, which create entanglement.
\autoref{fig:vqc_detail} visualises this gate sequence for one encoding and one variational layer.
This layer structure is chosen as it is seen as highly expressive throughout the literature~\cite{kandala17, skolik22}.
We considered two types of circuits:
In the first, we apply DRU and repeat the encoding pattern several times, which can increase quantum expressivity~\cite{perez20}.
In the second, we omit the input encoding part after each input feature is present in the circuit once, that is, we do not apply DRU, resulting in a flatter circuit.
Both variants are evaluated empirically in \autoref{sec:exp-qrl}.

\section{Experiments}
\label{sec:exp}

We commence with the experimental setup
(fully reproducible with our \repro), followed by the training results for quantum-based versions of \rj in \autoref{sec:exp-qrl}.

\subsection{Experimental Setup}
\label{subsec:setup}

\subsubsection{Training and Test Data}
Following the approaches presented in \autoref{sec:rel-work} that evaluate their methods using various industrial benchmark datasets~\cite{leis15, poess02, poess10},
for classical \rj, we used the 113 queries from the join order benchmark (JOB)~\cite{leis15}.
As we lack access to a sufficiently large quantum machine to process data for all queries in the JOB, we concentrate on training with four relations per query.
Since the JOB only provides three queries with four relations, we generate new queries based on subplans to enlarge the dataset, following Krishnan~\etal~\cite{krishnan18}.
However, instead of obtaining subplans from the traditional optimiser, we rely on a single \rj training run, generating over \numprint{12000} subqueries, from which we randomly select \numprint{497} that join four relations,
and combine them with three JOB queries.
To the resulting dataset of size \numprint{500}, we apply a ten-fold cross-validation scheme~\cite{fushiki2011}, whereby the dataset is split into ten distinct parts.
Each part is excluded from the training set once to be utilised for testing, leading to ten different train-test-splits.

\subsubsection{Python Libraries}
Since the original source code for \rj is not available, and other implementations for solving the JO problem by the means of RL~\cite{rtos_impl, yang22, marcus21} utilise different RL methods~\cite{mnih13}, and a different encoding for states and actions~\cite{marcus21, yang22} we modified and fine-tuned a third-party replication~\cite{rejoin_impl} based on the descriptions in Ref.~\cite{marcus18} in collaboration with one of the original authors using the Python machine learning library \emph{Tensorflow}~\cite{abadi15} for the machine learning specific parts.
For the quantum specific parts of our experiments, we additionally utilised the quantum frameworks \emph{Tensorflow Quantum}~\cite{broughton21} to simulate ideal quantum systems and \emph{Qiskit}~\cite{qiskit} to simulate noisy systems.
Given the lack of capable quantum machines, we rely on simulations.
The implementation can be found in our \repro.

\subsubsection{Classical Baseline Replication}
We were able to successfully replicate \rj, despite some minor deviations from the findings in Ref.~\cite{marcus18}, which could possibly attained to differing hyperparameters or settings that were not specified in the original study.
To further enhance the outcomes of our replication, and to allow for the reduced encoding described in \autoref{sec:mod-reduced}, we combined methods from other RL approaches for JO~\cite{krishnan18, xiang20} to improve the learning convergence in cost training.
For more information on the classical replication and the baseline modification, the reader is referred to the supplementary material in the \repro.

\subsubsection{PPO Models}
We consider the following configurations:

\paragraph{Classical Model}
As baseline, we use a classical NN with two hidden layers (128 units each) for actor and critic.

\paragraph{Quantum Model---Single-Step~\cite{winker23}}
This model uses one qubit per relation in the query, resulting in 4 qubits for our dataset. For each relation in the query, the ID of the relation is encoded with an $\hat{R}_x$ gate and the combined selectivity of all filters on the relation is encoded with a $\hat{R}_y$ gate. As there are at most 15 possible join orders for 4 relations, $2^4$ quantum states are enough to have a state for each join order.

\paragraph{Quantum Models---Multi-Step}
We consider three configurations: (a)
\emph{Q-Critic}, where a VQC is employed as the critic part of PPO,
and a classical NN with the same dimensions as for the classical 
model serves as the actor; (b) \emph{Q-Actor} with a VQC
as actor in PPO, and classical critic; (c) \emph{Fully Quantum}
with VQCs for actor and critic. All quantum models use classical post-processing layer (see \autoref{subsec:dec}).

\subsubsection{Data Re-Uploading (DRU) Setup}

For each quantum model, we evaluate setups with and without DRU.
The version utilising DRU employs 2--5 repetitions of the gates necessary to encode all input features once.
With four relations this results in 8, 12, 16 and 20 variational layers for the multi-step QRL approach.
To ensure a fair comparison with the single-step QML approach, we repeat the input features, consisting of indices and selectivities, every four variational layers for the configurations with single-step QML and DRU.
This results in the same number of input repetitions and variational layers as for the multi-step QRL approach.
For the second configurations without DRU, we use the same number of variational layers and introduce an additional experiment with four layers to encode every input feature once without extra variational layers for multi-step QRL.
Analogously, the configurations for single-step QML and without DRU encode the input features once followed by the respective number of variational layers.

\subsubsection{Training and Evaluation}
While incorporating noise during training, whether through direct execution on real QPUs or via noisy simulations utilising snapshots of actual devices, provides the most accurate assessment of our approach's performance on present or near-term quantum hardware, the computational demands of noisy simulation, particularly for large input sizes during optimisation, are substantial.
Given these constraints, a complete training iteration exceeds the scope of this study.
Nonetheless, to quantify the adverse effects of noise, we assess models trained in an ideal simulation in a noisy environment using the same test sets from the ten-fold cross-validation.
Specifically, we introduce depolarising errors~\cite{nielsen10}, a prevalent error type in noisy simulations, with a predetermined probability applied to each gate within the models utilising a quantum actor (\ie, \emph{Q-Actor} and \emph{Fully Quantum}).
For this probability, we select values ranging from 1\% to 5\%, representing upper bounds of gate errors, to which current QPUs are prone~\cite{greiwe23}.
The findings from our noisy evaluation are detailed in \ref{sec:noisy-validation}.

\subsection{Experimental Results}
\label{sec:exp-qrl}

\subsubsection{Training Results from Ideal Simulations}

\begin{figure}[htbp]
  \includegraphics{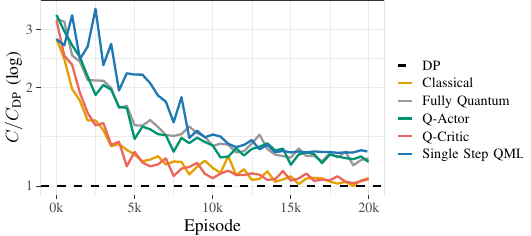}\vspace*{-0.5em}
  \caption{Relative cost median during training. For the methods involving a quantum part the models with DRU and 20 variational layers are depicted.}
  \label{fig:rejoin-q-rels4}
\end{figure}

As shown in \autoref{fig:rejoin-q-rels4}, the classical model can achieve the median for optimal join orders after sampling roughly \numprint{8000} queries (episodes), while Q-Critic delivers comparable results.
Since the single-step approach surpasses conventional JO heuristics~\cite{winker23, calikyilmaz23} when trained on query execution times (\ie, true cost), it can be
regarded as quantum baseline.  We either outperform or match it in all three QRL variants.
Specifically, the Q-Critic configuration can achieve up to 17\% lower median costs than single-step QML.
This implies that although the configurations that employ a VQC as an actor, as well as the single-step QML method, do not achieve an optimal cost median during training, the QRL approaches are competitive with established classical heuristics, assuming that careful hyperparameter tuning and incorporating true costs leads to better join orders.
Since our focus is on the specific implications for quantum computing, and as training on a cost model may not necessarily translate to actual query execution times, we consider costs as performance indicator, following Refs~\cite{marcus18, krishnan18, xiang20}.

Our results suggest that as the classical component of computation increases, the quality of results improves.
This finding appears to contradict claims for quantum advantage in QML literature~\cite{havl19, huang21, liu21}.
However, it aligns with a recent observation by Bowles~\etal~\cite{bowles24} who conducted benchmarks across various QML configurations and noted that models with a substantial portion of classical parameters often outperform those with a higher quantum component.
Understanding the dynamic between classical and quantum methods remains an important future challenge.

\begin{figure}[htbp]
  \includegraphics{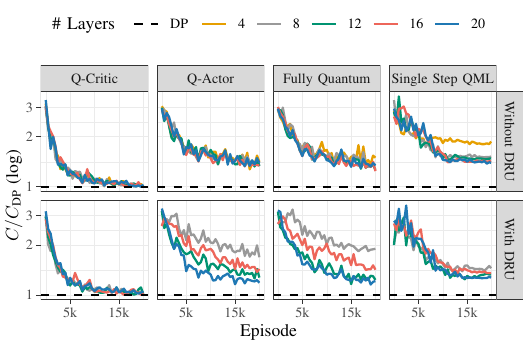}\vspace*{-0.5em}
  \caption{Relative cost median during training.}
  \label{fig:rejoin-q-rels4-layers}
\end{figure}

As illustrated in \autoref{fig:rejoin-q-rels4-layers}, the quantity of variational layers impacts configurations with a higher proportion of VQC parameters (Q-Actor, Single-Step QML and Fully Quantum QRL), especially with DRU.
We observe, consistent with findings in the literature~\cite{winker23, skolik22, jerbi21, dilip22} that more layers lead to lower costs.
In all other instances, optimal training convergence is attainable  with fewer variational layers, which translates to fewer parameters and shallower circuits.

\subsubsection{Evaluation Results from Noisy Simulations}
\label{sec:noisy-validation}

\begin{figure}[htbp]
  \includegraphics{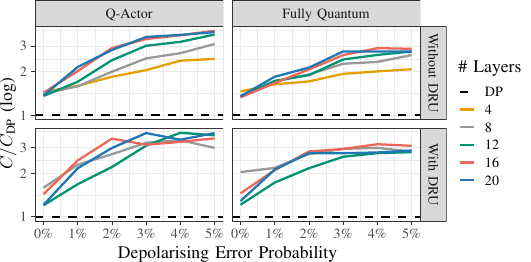}\vspace*{-0.5em}
  \caption{Relative cost median after training with different noise probabilities.}
  \label{fig:noisy-validation-median}
\end{figure}

\begin{figure}[htbp]
  \includegraphics{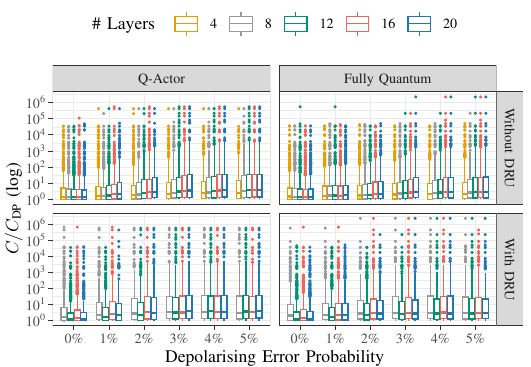}\vspace*{-0.5em}
  \caption{Relative costs after training with different noise probabilities.}
  \label{fig:noisy-validation}
\end{figure}

The configurations utilising a quantum actor (\ie, \emph{Q-Actor} and \emph{Fully Quantum}) demonstrate the capability to achieve nearly optimal results conducted on ideally simulated QPUs.
However, when incorporating gate errors, the performance of quantum models tends to deteriorate.
\autoref{fig:noisy-validation-median} shows that the median relative cost increases almost linearly across all configurations, with steeper increases observed for deeper circuits---those with more layers and 
DRU--—, which inherently present more opportunities for errors.
While this observation is sobering, it aligns with the expectation that models trained in a ideal environment may struggle when confronted with noise.
Other studies~\cite{skolik22, borras23} suggest that incorporating noise during training, coupled with hyperparameter tuning tailored to such noise models, can yield successful outcomes even in the presence of noise.
The exploration of noise's impact during training on the JO problem could be deferred to future investigations.
However, as shown in \autoref{fig:noisy-validation}, a comprehensive examination of results reveals that significant outliers persist, even in ideal and classical scenarios, indicating that while median performance appears reasonable, pronounced instabilities persist within the (Q)RL approach to the JO problem, necessitating further theoretical and empirical investigation of the methods itself.

\section{Evaluation}
\label{sec:discussion}

While the quantum models may not outperform classical models in terms of cost efficiency post-training, other factors are pertinent to assess the effectiveness of QML methods for JO.
This section discusses the influence of the circuit dimensionality on overall trainability, examining the number of parameters and scalability of our approaches compared to alternative quantum approaches for the JO problem.

\subsection{Parameter Efficiency}

\begin{figure}[htbp]
  \includegraphics{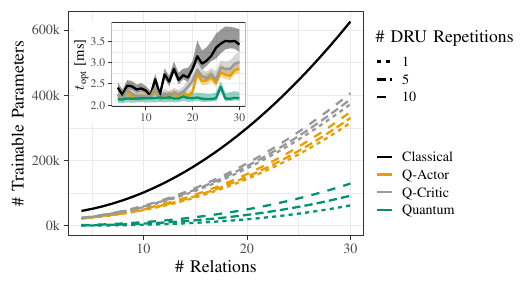}\vspace*{-1em}
  \caption{Number of parameters for classical and quantum methods. The pure 
  quantum case requires substantially less parameters than the classical baseline,
  which reduces optimisation complexity. Partial quantum variants (Q-actor, Q-critic)
  are less parsimonious, yet retain advantages against the baseline. The inset shows the optimisation time of the Adam~\cite{kingma17} optimiser, to apply gradients to the parameters. Lines show the median of \numprint{1000} measurements for each configuration; shaded areas depict first and third quartiles. Note that gradient calculation is excluded from our measurements.}
  \label{fig:num_params}
\end{figure}

\autoref{fig:num_params} shows the total number of parameters 
(variational and classical) dependent on the number of relations in 
a query for the different methods.
We considered the VQC structure with the dimensions described in \autoref{subsec:dimensions}, including the classical post-processing layer and a fully-connected NN with two hidden layers and a constant hidden dimension of 128\footnote{Typically, the number of hidden units grows with input space~\cite{sheela13}, so 
the number of parameters for the classical model gives a lower bound.}.
While the classical baseline achieves lower and more stable costs, the QRL variants require fewer parameters.
Considering the Q-Critic configuration, which achieves costs 
comparable to the baseline, we found that 47\% less parameters suffice for four relations using 20 variational layers, that is, five DRU repetitions, and about 38\% less parameters for 30 relations.

The corresponding run-times for the \emph{Adam optimiser}~\cite{kingma17} are shown in the inset of \autoref{fig:num_params}.
We did not consider the time taken to calculate the gradients in our optimisation time measurements, as (1) gradient calculation or estimation methods for VQCs are still an ongoing area of research~\cite{schuld18, wierichs22, gilyen19} and (2) our experiments are conducted on simulators instead of real QPUs, so the execution times may differ significantly.
The parameter-shift rule~\cite{schuld18}, commonly used with VQCs, is computationally and necessitates two circuit executions per shot and parameter. Optimised techniques for gradient calculations have appeared~\cite{periyasamy:2024,Stokes_2020,bittel2022fast,Spall92},
similar to classical ML over the past decades~\cite{sarker21}. Yet, a comprehensive
evaluation is beyond the scope of this paper.
Based on our measurements, it is possible to achieve up to 12\% improvement in median optimisation time for the Q-Critic configuration with one DRU repetition, in comparison to the classical model per optimisation step for 30 relations.
As ML methods update parameters over multiple thousand iterations, this could significantly impact overall training time.

\subsection{Scalability of Quantum Approaches for Join Ordering}
\label{sec:scalability}

\begin{figure}[htbp]
  \input{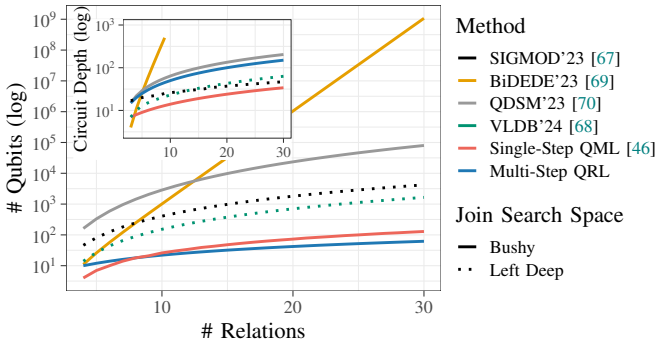}\vspace*{-2em}
  \caption{Number of qubits and circuit depth required to encode the JO problem for different quantum optimisation strategies.}
  \label{fig:num_qubits}
\end{figure}

As outlined in \autoref{sec:rel-work}, other quantum-based techniques address the JO problem.
The number of qubits necessary to encode up to 30 relations for each of these strategies is depicted in \autoref{fig:num_qubits}.
Refs.~\cite{nayak23, schoenberger23, schoenberger23:qdsm, schoenberger24} aim to solve a specific class of problems, namely quadratic unconstrained binary optimisation (QUBO) problems, where the number of qubits required depends on the QUBO formulation.
In contrast, QML approaches provide greater flexibility in the utilisation of qubits and circuit depth.
As shown in the figure, both proposed QML approaches, single-step QML and multi-step QRL, are more efficiently in terms of qubit numbers, compared to the QUBO approaches.
Furthermore, circuit depth is a widely accepted quantum runtime proxy, for which we provide bounds%
\footnote{Bounds are based on circuit depth for one data uploading block (QML approaches) and lower bounds on the circuit depth for the respective QAOA~\cite{farhi14} circuit with \(p=1\) (QUBO-based approaches).
We consider the maximum number of entangling gates/quadratic terms that act on two qubits, and gates required for initialisation and mixer Hamiltonian.}
in~\autoref{fig:num_qubits}.
QRL generally requires only low circuit depth, comparable to the 
QUBO approach for bushy joins presented in
Ref.~\cite{schoenberger23:qdsm}.
However, not unlike with classical machine learning~\cite{goodfellow16}, while substantial progress with
understanding capabilities of VQCs has been made~\cite{landman22}, the learning dynamics based on the circuit dimension and theoretical underpinnings are not yet fully understood~\cite{schuld22} and require further empirical and theoretical evaluation.

\section{Discussion and Outlook}
\label{sec:concl}
We introduced a quantum reinforcement learning based approach to solve the long-standing, seminal join order problem, and replicated a
classical reinforcement learning approach as suitable baseline to compare against the state of the art.
In a systematic and comprehensive evaluation based on numerical simulation of quantum systems, we found that our approach at least matches classical performance in terms of result quality, which is not universally observed throughout the literature~\cite{meyer22} for quantum algorithms.
Apart from significantly reducing the input feature space of the classical baseline, we could show that substantially fewer trainable parameters are required, which is likely
rooted in enhanced quantum expressivity.
We believe the resulting reduction in classical optimisation efforts particularly benefits two scenarios: (a) Frequently changing data characteristics that necessitate continuous re-computation of join orders, and (b) low response latency requirements. Both appear in important commercial settings like stream data processing and high-frequency operation~\cite{Dean:2013}. 

We also showed that our approach improves upon the scalability of existing quantum-RL solutions by nearly ten orders of magnitude in terms of qubit count. Given that this is the most scarce resource in current and future QPUs, we believe this is an important step towards practical utility. 

Current NISQ capabilities prevents us from enjoying practical 
advantages right now.
The limitations might, however, be circumvented even prior to the arrival of fully error-corrected hardware that is capable of delivering the behaviour predicted in our simulations by using custom-designed hardware.
Additionally, it has recently been observed that the JO problem on 
quantum-inspired hardware can outperform established approaches~\cite{schoenberger24}.
Similar observations could generalise to other types of hardware, potentially applicable to the domain of variational algorithms or machine learning that our approach is based on.
Finally, progress in the foundational understanding of QML could improve performance
using more sophisticated quantum baseline methods, or data encoding strategies.

\newcommand{\MF}{\censor{MF}\xspace}
\newcommand{\WM}{\censor{WM}\xspace}
\newcommand{\TW}{\censor{TW}\xspace}
\newcommand{\SG}{\censor{SG}\xspace}
\newcommand{\programme}{\blackout{German Federal Ministry of
Education and Research (BMBF), funding program \enquote{Quantum Technologies---from
Basic Research to Market}}}
\newcommand{\grantsoth}{\censor{\#13N15647 and \#13NI6092}}
\newcommand{\grantuzl}{\censor{\#13N16090}}
\newcommand{\hta}{\blackout{High-Tech Agenda Bavaria}}

\begin{small}
\noindent\textbf{Acknowledgements} \MF, \TW, \SG and \WM were supported by
the \programme, grants \grantsoth\ (\MF and \WM), and \grantuzl\ (\TW and \SG).
\WM acknowledges support by the \hta.
\end{small} 

\FloatBarrier\clearpage

\printbibliography

\end{document}